\documentclass[5p]{elsarticle}

\usepackage{graphicx}
\usepackage{dcolumn}
\usepackage{bm}
\usepackage{longtable}
\usepackage{float}
\usepackage[fleqn]{amsmath}

\makeatletter
\def\ps@pprintTitle{%
 \let\@oddhead\@empty
 \let\@evenhead\@empty
 \def\@oddfoot{}%
 \let\@evenfoot\@oddfoot}
\makeatother

\usepackage{lineno,hyperref}
\modulolinenumbers[5]

\begin{document}

\begin{frontmatter}

\title{Ultrafast demagnetization at high temperatures}

\author{F. Hoveyda}

\author{E. Hohenstein}

\author{R. Judge}

\author{S. Smadici \corref{cor1}}


\address{Department of Physics and Astronomy, University of Louisville, KY 40292, USA}%

\begin{abstract}
Time-resolved pump-probe measurements were made at variable heat accumulation in Co/Pd superlattices. Heat accumulation  increases the baseline temperature and decreases the equilibrium magnetization. Transient ultrafast demagnetization first develops with higher fluence in parallel with strong equilibrium thermal spin fluctuations. The ultrafast demagnetization is then gradually removed as the equilibrium temperature approaches the Curie temperature. The transient magnetization time-dependence is fit well with the spin-flip scattering model.
\end{abstract}


\end{frontmatter}


\section{Introduction}

Ultrafast demagnetization (UDM) is a laser-induced  transient state, during which magnetization evolves several orders of magnitude faster than magnetization precession. It demonstrated the magnetic materials potential for practical applications at high frequencies, at which the macrospin approximation is no longer valid. The transient magnetization time-dependence $M(\tau)$ has been measured in many materials, for instance in Ni and Gd~\cite{2010Koopmans,2017Bierbrauer,2017Bobowski}, FeRh~\cite{2014Gunther} and FePt compounds~\cite{2014Mendil}, and ferromagnetic Co/Pt superlattices~\cite{2014Kuiper,2017Medapalli}. Stochastic models~\cite{2008Kazantseva,2009Kazantseva}, scattering off impurities~\cite{2005Koopmans}, electron-phonon spin-flip scattering~\cite{2010Koopmans,2014Nieves}, non-local spin and electron diffusion models~\cite{2010Battiato,2016Salvatella,2016Bergeard,2017Eschenlohr,2015Schmising}, have been considered.

UDM may be viewed as a laser-induced non-equilibrium spin fluctuation added to equilibrium thermal spin fluctuations. Measurements with equilibrium sample temperature increases up to $480~K$ and large UDM transients from strong pulses were consistent with predictions of spin-flip scattering models~\cite{2012Roth}. Different conditions are present when thermal spin fluctuations are larger than UDM transients. In contrast to experiments at low repetition rates on thick and thermally-conducting films, heat accumulation from high repetition rate lasers and thin samples on thermally-insulating substrates enable local variation of the equilibrium temperature up to the Curie temperature~\cite{2017F-JAP}. In addition, a smaller transient signal compared to previous UDM experiments is obtained with smaller pulse energies, associated with the high repetition rate.

In this work, time-resolved pump-probe measurements were made on Co/Pd superlattices at variable heat accumulation temperatures, when equilibrium thermal spin fluctuations dominate the ultrafast demagnetization transients. Partial, then complete demagnetization, from transient UDM and equilibrium heat accumulation are observed. Thermal spin fluctuations gradually reduce the average magnetization and the transient UDM induced by the pump pulse. Measurements confirm that full thermal demagnetization, at a Curie temperature $T_{C}=610~K$, is obtained before all-optical switching in Co/Pd superlattices in our experimental conditions. The spin-flip scattering model is applied to fit the transient demagnetization measurements. 

\section{Experiments}

\subsection{Setup}

Ferromagnetic $\rm [Co/Pd]_{4}$ superlattices were examined, in which cumulative all-optical switching (AOS) was observed with linearly-polarized light~\cite{2017F}. The samples were $h=4.1~nm$ thick with perpendicular magnetic anisotropy.

The pump-probe two-frequency setup has a non-collinear geometry, with measurements in transmission at normal incidence (figure 1). The linearly-polarized 800 nm pump and 400 nm probe beams, with pulses of $\tau_{p}=190~fs$ duration, were focused to stationary $w_{0}=125~\mu m$ and $w_{1}=80~\mu m$ spots, respectively, and the delay between the two pulse sequences scanned with a translation stage.

The sample magnetization was re-initialized between pump pulses with a constant field $|B|=300~G$ from two water-cooled coils. The relatively strong damping $\alpha\approx 0.1$ in Co/Pd~\cite{2011Liu,2011Pal} insures that the magnetization is stable within the $12.5~ns$ time interval between pulses. This allows measuring transient processes with the same initial and final states. Measuring the AOS time dependence, with different initial and final states, requires that magnetization be reset between the pulses with a field pulsed at the TiS laser repetition rate ($\rm 80~MHz$) and cannot be currently done in our setup.

A balanced photodiode lock-in detection at the pump beam chopping frequency $f=2.069~kHz$ has been applied. The lock-in resultant $R$ voltage dependence on delay $\tau$ for different applied fields $B=\pm 300~\rm G$ was separated into two components, $S(\tau)=\frac{1}{2}\Big( R(-B)+R(+B)\Big)$ and $A(\tau)=\frac{1}{2}\Big( R(-B)-R(B)\Big)$, symmetric and anti-symmetric in $B$, respectively. Intensity and polarization variations arise from temperature and birefringence transients. A configuration with a polarizer and analyzer near crossing minimizes non-magnetic contributions to the anti-symmetric component $A(\tau)$~\cite{2003Koopmans-book}. Measurements away from crossing configuration resulted in a featureless $A(\tau)$.

\subsection{Results}

\begin{figure}
\centering\rotatebox{0}{\includegraphics[scale=0.35]{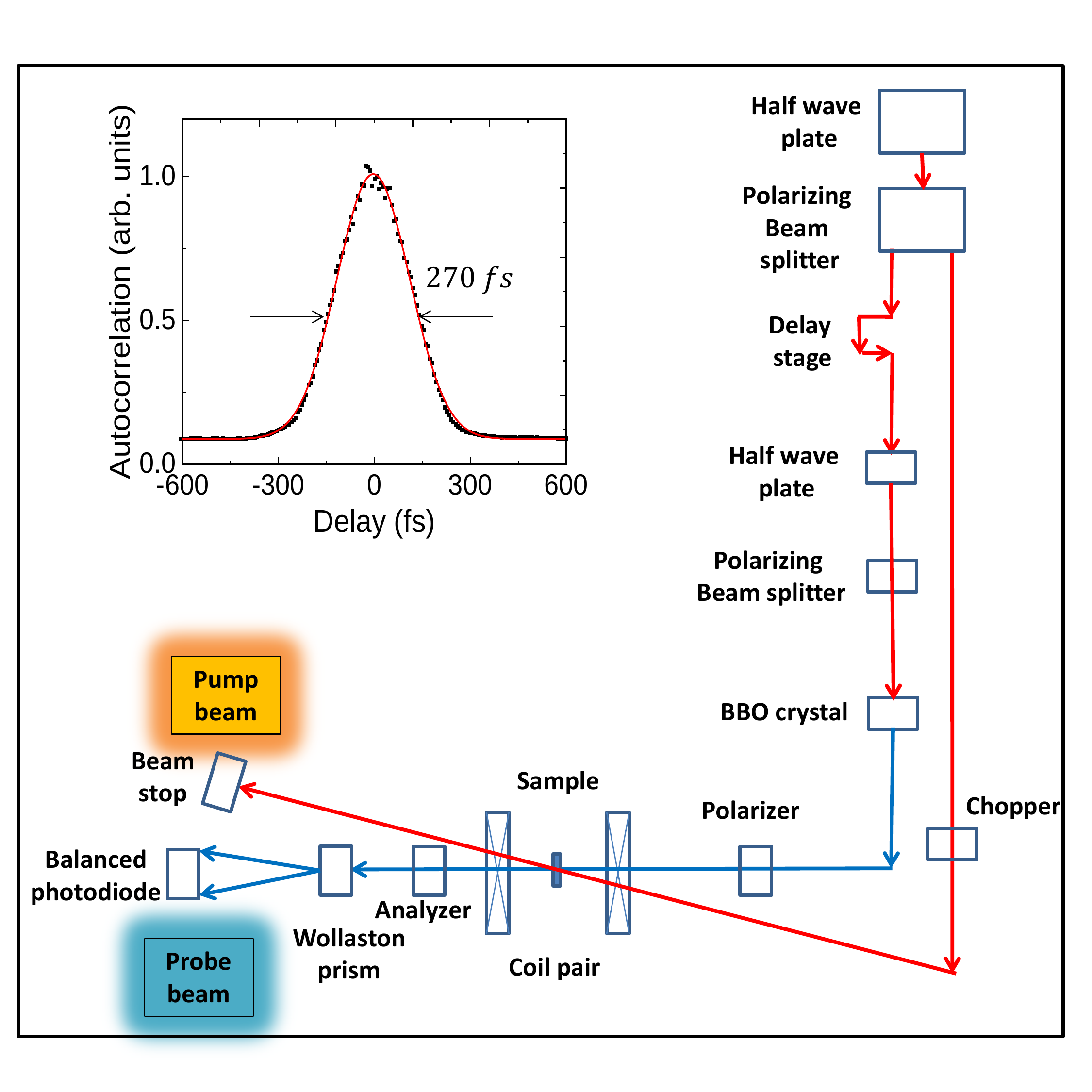}}
\caption{\label{fig:Figure1} Sketch of the experimental setup. Inset: autocorrelation of $\tau_{p}=190~fs$ pump pulses at sample location.}
\end{figure}

The symmetric component $S(\tau)$, obtained from measurements at different pump beam powers (figure 2), shows an overall offset $S_{eq}$ due to heat accumulation from multiple pulses~\cite{2017F-JAP}, and a prominent peak, similar to results for Co films~\cite{2005Bigot}, with a step $S_{step}$ from one-pulse transients (figure 3). A Gaussian with a step function $S(\tau)=S_{1}G(\tau-\tau_{C})+S_{2}\theta(\tau-\tau_{C})+S_{eq}$ fits the experimental results well (figure 3).

The antisymmetric component $A(\tau)$ (figure 4(a)) time-dependence corresponds to type I UDM~\cite{2010Koopmans}, with demagnetization time smaller than equilibration time $\tau_{M}<\tau_{E}$, similar to Co and Co/Pt~\cite{2014Kuiper}, and consistent with measurements of UDM in Co/Pd with XMCD~\cite{2010Boeglin}, XRMS~\cite{2016Vodungbo}, X-ray Fourier transform holography~\cite{2014Schmising}, and optical Kerr Effect~\cite{2015Schmising}. A rate equation double-exponential fit was applied to quantify the overall variations with fluence

\begin{eqnarray}
A(\tau)= \Big( A_{1} e^{-\frac{\tau}{\tau_{E}}}- A_{2}  e^{-\frac{\tau}{\tau_{M}}}\Big)\theta(\tau)+ A_{eq}.
\end{eqnarray}

\noindent where $\theta(\tau)$ is the unit step-function. The measurements were well fit with $\tau_{M}=0.25~ps$, $\tau_{E}= 1.3~ps$, and $A_{1}=A_{2}$. $\tau_{M,E}$ do not show a discernible dependence on fluence, unlike the results for $\tau_{E}$ in Ref.~\cite{2012Vodungbo}. A more detailed fit is done in section 3.

\begin{figure}
\centering\rotatebox{0}{\includegraphics[scale=0.4]{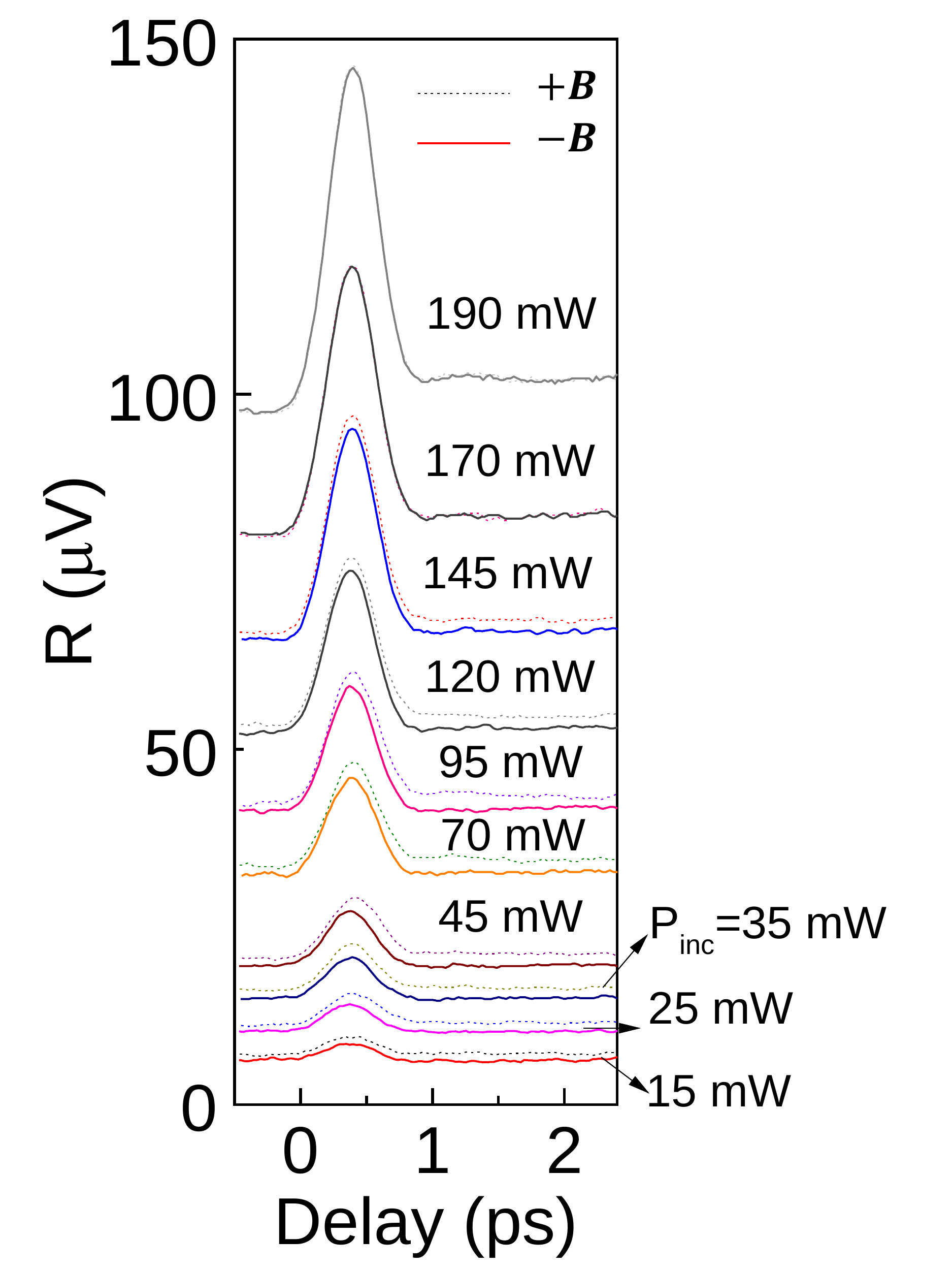}}
\caption{\label{fig:Figure1} $R(\tau)$ for different constant magnetic fields and incident pump beam powers.}
\end{figure}

The UDM amplitude $A_{1,2}$ has a non-monotonic dependence on power (figure 4(b)). It first increases with fluence, as observed before. In contrast to previous measurements, heat accumulation is significant and the amplitude decreases with further increases of fluence.

Heat accumulation temperature $T_{acc}$ cannot be neglected for thin samples and high-repetition rate lasers, with multiple pulses incident on the same area within the heat diffusion time $\frac{w_{0}^{2}}{4D}$. Complete demagnetization by thermal spin fluctuations at high $T_{acc}$ is obtained at an incident pump beam power $P_{inc}=170~mW$ which, from measurements of reflected and transmitted beam powers, corresponds to an absorbed power $P_{abs}=40~mW$. The increase of equilibrium temperature above the room temperature at the center of the beam is $T_{acc,max}=310~K$ for $P_{abs}=40~mW, h=4.1~nm, w_{0}=125~\mu m$, an interface conductance $G>10^{6}~W/m^{2}K$ and the same thermal parameters as in Ref.~\cite{2017F-JAP}. This gives a Curie temperature $T_{C}=610~K$, consistent with results in similar samples of $T_{C}=800~K$ for Co/Pd~\cite{2016Chen} and $T_{C}=600~K$ for Co/Pt~\cite{2014Kuiper}. The $A(\tau)$ plots have been stacked along a Brillouin $m(T)= M(T)/M_{0}$ function for spin $1/2$, normalized to the magnetization $M_{0}$ at $T=0$, according to the heat accumulation temperature calculated at the center of the beam (figure 4(a)).

\begin{figure}
\centering\rotatebox{0}{\includegraphics[scale=0.4]{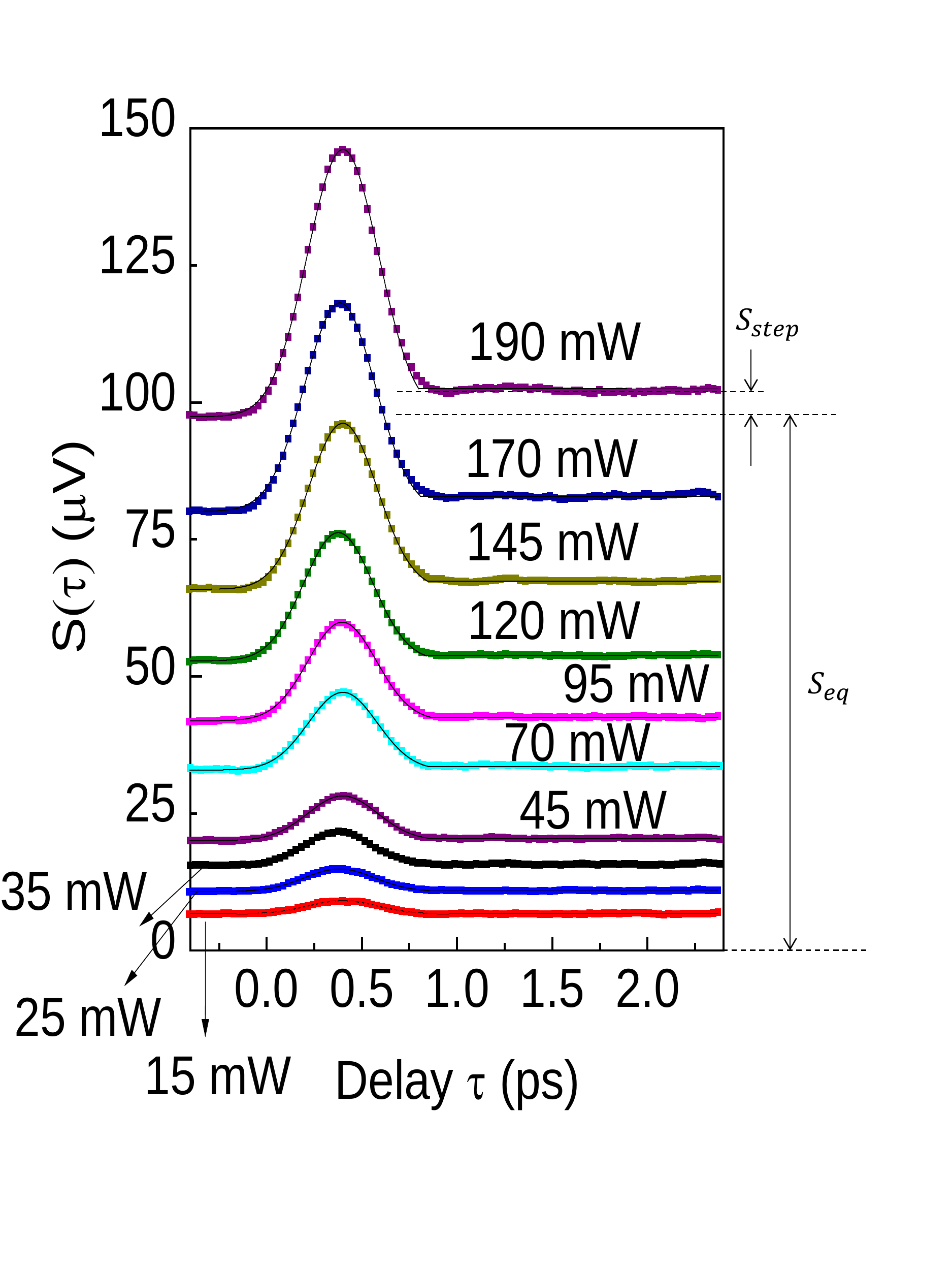}}
\caption{\label{fig:Figure1} $S(\tau)$ at different power and Gaussian with step fit.}
\end{figure}

\begin{figure}
\centering\rotatebox{0}{\includegraphics[scale=0.4]{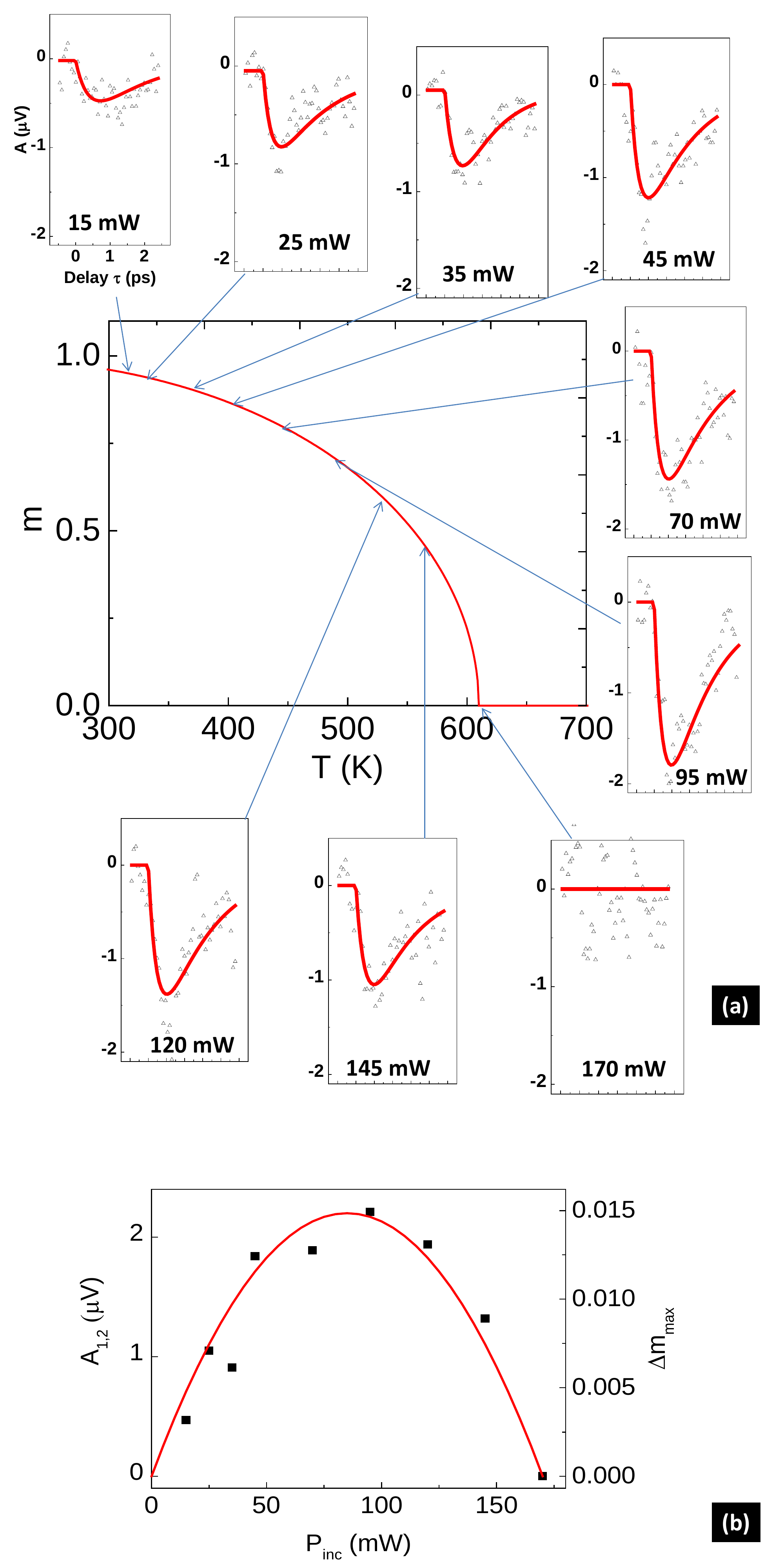}}
\caption{\label{fig:Figure1} (a) Double-exponential fit of the transient $A(\tau)-A_{eq}$ at different power, showing the interference of thermal spin fluctuations with UDM at high heat accumulation temperatures. (b) $A_{1,2}$ and the maximum transient demagnetization $\Delta m_{max}$ vs. the incident beam power. The maximum transient demagnetization, increasing up to $\sim 1.5~\%$, is much smaller than the equilibrium thermal demagnetization, ranging from $4~\%$ to $100~\%$ (panel (a)).}
\end{figure}

Measurements confirm that demagnetized states are induced in Co/Pd before AOS. Specifically, the time-resolved measurements were made with a $w_{0}=125~\mu m$ diameter stationary beam. The AOS measurements were made with a $w_{0}'=50~\mu m$ diameter moving beam. The moving beam introduces a speed-dependent reduction factor of $f=0.75$ and $0.3$ in $T_{acc}$ compared to a stationary beam, for $1~mm/s$ and $10~mm/s$ speeds, respectively~\cite{2017F-JAP}. Therefore, the $170~ mW$ power for full demagnetization in time-resolved experiments (figure 4(a)) corresponds to a $(\frac{w_{0}'}{w_{0}})^{2}\times \frac{170~ mW}{f}$ power in the AOS experiments, or $36~ mW$ and $90~ mW$ for $1~ mm/s$ and $10~ mm/s$ speeds, respectively. This is less than the power required for AOS~\cite{2017F}. A fully demagnetized state is a precursor to cumulative AOS in Co/Pd. The demagnetized state can then evolve to a reversed state~\cite{2018F}.

\section{Discussion}

The equilibrium lattice temperature was varied over a wide range in our experimental conditions, with a relatively small transient lattice temperature induced by each low-energy pulse. The magnetization dynamics, determined by energy and angular momentum dissipation into the lattice, has been examined in detail in the spin-flip scattering model with very good fits to the measurements~\cite{2010Koopmans}.

The measurements will be fit with the spin-flip model. An unexplored connection between UDM and the spin-boson model is applied first to obtain the spin-flip model functional dependence. 

\subsection{Stochastic models of longitudinal magnetization dynamics}

The time-evolution of a system out of equilibrium with its environment can be examined with stochastic methods. For instance, fluctuations of a superparamagnetic particle macrospin have been investigated with fluctuating magnetic fields $\zeta_{i}$ added to the Landau-Lifshitz-Gilbert equation~\cite{1963Brown}. This gives a Langevin equation for each macrospin, a Fokker -- Planck equation for their probability distribution with a diffusion in spin orientations, and a Bloch-type equation for the average magnetization~\cite{1963Brown,1970Kubo,1998Garcia}.

This method has been applied to laser-induced ultrafast magnetization dynamics of ferro-~\cite{2009Kazantseva} and ferrimagnetic~\cite{2012Atxitia} films and particles, where the macrospin is fragmented in microscopic spins $S_{i}$ by the strong pump pulse. Each spin evolves according to a Langevin equation with stochastic fields $\zeta_{i}$~\cite{2008Kazantseva}. This gives an equation for the average magnetization $M$, with an additional longitudinal damping term added to the standard precession and transverse damping. Keeping only the longitudinal term at the UDM timescale, the fragmentation and gradual reassembly during UDM of the normalized magnetization $m_{z}=M_{z}/M_{0}$ is given by

\begin{eqnarray}
\frac{dm_{z}}{dt}=\frac{\gamma \alpha _{||}}{2\tilde{\chi}_{||}} \Bigg(1-\frac{m^{2}}{m_{eq}^{2}}\Bigg)m_{z} \approx -\frac{\gamma \alpha _{||}}{\tilde{\chi}_{||}} \Bigg(\frac{m-m_{eq}}{m_{eq}}\Bigg)m_{z}
\end{eqnarray}

\noindent where $T < T_{C}$, $\gamma$ is the gyromagnetic factor, $\alpha_{||}$ the longitudinal damping factor, and $\tilde{\chi}_{||}$ the longitudinal magnetic susceptibility~\cite{2009Kazantseva,1970Kubo,1997Garanin,2014Evans,2015Hinzke,2017Atxitia}. In equilibrium, $1/\tilde{\chi}_{eq}=kT/\sigma_{M}^{2}=\sigma_{B}^{2}M_{0}/kT$, where $\sigma_{M}=M_{0}\sqrt{\langle m_{z}^{2}\rangle - \langle m_{z}\rangle^{2}}$ and $\sigma_{B}=\sqrt{\langle \zeta_{i}\zeta_{j}\rangle}$ are the standard deviations of magnetization and fluctuating fields with $\langle \zeta_{i} \rangle=0$.

The rate is proportional to $m-m_{eq}$ for small magnetization amplitude variations $m\approx m_{eq}$. Equal scattering rates for transitions in opposite directions gives a zero net rate in detailed balance equilibrium. The net rate is proportional to the difference $(m-m_{eq})$ for small deviations from equilibrium. The overall factor $(m^{2}-m_{eq}^{2})$ can also be obtained with the Landau model free energy near $T_{C}$~\cite{2009Kazantseva,1997Garanin}. The equilibrium magnetization is $m_{eq}(t) = \rm{tanh}\Big(\frac{\Delta}{2kT_{e}}\Big)$, where $\Delta$ is the energy splitting between up- and down-spin states, when spin fluctuations are averaged  with the mean-field approximation~\cite{2008Kazantseva,2009Kazantseva}. The pump pulse modifies $\Delta$ and $T_{e}$. In longitudinal dynamics $m$ follows a time-dependent magnetization amplitude $m_{eq}(t)$.

\begin{figure}
\centering\rotatebox{0}{\includegraphics[scale=0.4]{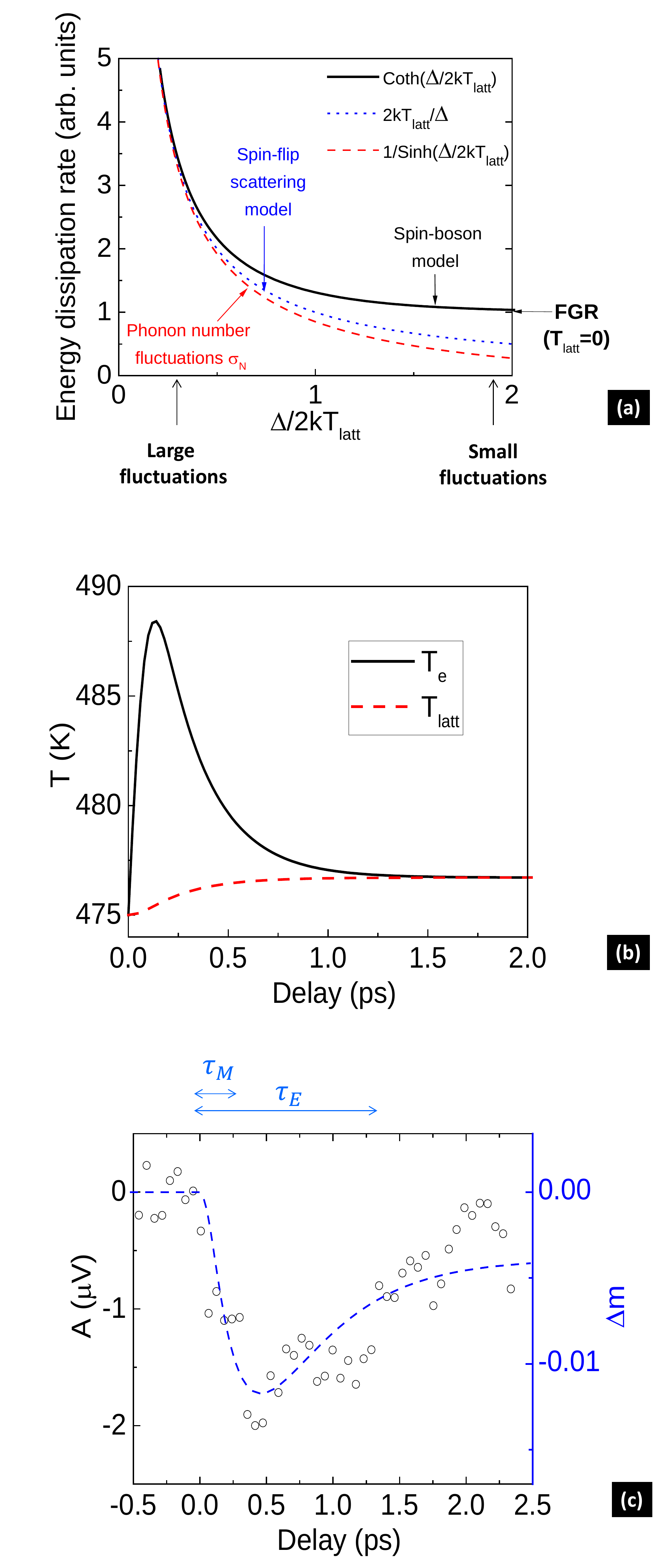}}
\caption{\label{fig:Figure1} (a) Energy dissipation rates at different $\Delta/(2kT_{latt})$. (b) Electron $T_{e}$ and lattice $T_{latt}$ temperature transients, for $P_{inc}=95~mW$, $T_{acc}=175~K$, $h=4.1~nm$ and $T_{C}=610~K$. Small pulse energies result in relatively small transient variations. (c) Measurements for $P_{inc}=95~mW$ (same as in figure 4) and calculations of magnetization variation.}
\end{figure}

\subsection{Energy dissipation into a harmonic oscillator bath}

A different approach to obtaining the time-evolution of a two-state system out of equilibrium with its environment has been developed with a spin-boson Hamiltonian~\cite{1987Leggett}. Remarkably, the energy dissipation rate $R$ of a two-state system with energy splitting $\Delta$ into an oscillator heat bath has a simple dependence on bath temperature

\begin{eqnarray}
R= 2 J(\omega)  \rm{coth}\Bigg(\frac{\Delta}{2kT_{latt}}\Bigg)
\end{eqnarray}

\noindent where $T_{latt}$ is the temperature and $J(\omega)$ is the spectral function of the bath (figure 5(a))~\cite{1974Nitzan}. For a bath with one optical phonon branch $J(\omega)=\pi \delta (\Delta-\omega)|G|^{2}$, where $\delta(\Delta-\omega)$ is the energy conservation factor and $G$ is the coupling constant between the two-state system and the heat bath. The Fermi Golden Rule (FGR) is obtained $R = 2\pi|G|^{2}\delta(\Delta-\omega)$ as $T\rightarrow 0$. The same temperature dependence results when adding phonon number fluctuations of a simple harmonic oscillator lattice mode $\sigma_{N}=\sqrt{\langle N^{2}\rangle - \langle N\rangle ^2}=\rm \frac{1}{sinh (\frac{\omega}{2kT_{latt}})}$ and the FGR contribution in quadrature $\rm \sqrt{\sigma_{N}^2+1} = coth\Big(\frac{\omega}{2kT_{latt}}\Big)$ (figure 5(a)).

Inserting $R$ into the result of the stochastic model gives

\begin{eqnarray}
\frac{dm_{z}}{dt}= A J(\omega=\Delta) \rm{coth}\Bigg(\frac{\Delta}{2kT_{latt}}\Bigg) \Biggl(1 - m ~\rm{coth}\Bigg(\frac{\Delta}{2kT_{e}}\Bigg)\Biggl) m_{z}
\end{eqnarray}

\noindent where $A$ is an overall factor. The magnetization time-dependence of the spin-flip scattering model~\cite{2010Koopmans} is obtained up to the overall factor $A$, since $\rm{coth} (\Delta/2kT_{latt}) \rightarrow 2kT_{latt}/\Delta$ in the limit $kT_{latt} \gg \omega=\Delta$. This occurs near the Curie temperature $T_{latt}\gg mT_{C}$ when $\Delta=2kT_{C}m$ of Ref.~\cite{2010Koopmans} is applied. Finding the overall factor $A$ requires a scattering model, as done for an acoustic phonon spectrum~\cite{2010Koopmans}.

A two-step procedure was followed to fit the measurements. First, heat accumulation was calculated, to give the baseline temperature $300~K+T_{acc}$. Then, transients were added with two-temperature rate models and longitudinal magnetization dynamics. The electron and lattice temperature time-dependence are shown in figure 5(b) for $P_{abs}=22~mW$, corresponding to an incident power $P_{inc}=95~mW$ and $E_{pulse,abs}=0.3~nJ$, with electron-phonon coupling $g_{ep}=75\times10^{16}~W/(m^{3}K)$ and an electronic specific heat coefficient $\gamma_{e}=400~J/(m^{3}K^{2})$. The approximate magnitudes can be estimated from energy conservation, $T_{e,max}\approx \sqrt{\frac{2E_{pulse,abs}}{\gamma_{e} hw_{0}^{2}}+T_{0}^{2}}\approx 499~K$, when neglecting the transfer of energy to the lattice over the duration of the pulse, where $T_{0}=475~K$ is the initial temperature, including the heat accumulation. Similarly, the lattice temperature step increase after one pulse is $T_{latt}\approx E_{pulse}/(C_{latt} hw_{0}^{2}) \approx 1.5~K$. The magnetization was calculated with the spin-flip model and is compared to measurements for $P_{inc}=95~mW$ in figure 5(c). The maximum transient demagnetization is $\approx 1~\%$, smaller than the $26~\%$ equilibrium demagnetization at this power (figure 4(a)).

The measured demagnetization $\langle \Delta m(\tau) \rangle$ is an average over probe beam area and measurement time at each delay $\tau$. Spin fluctuations are averaged with the mean-field approximation and the reduction of UDM amplitude at high pump power shows the increased thermal spin fluctuations $\sigma_{M}$ indirectly. Slower superparamagnetic particle equilibrium macrospin fluctuations have been measured~\cite{1997Wernsdorfer} and obtained numerically from the Langevin equation~\cite{1998Garcia}. X-ray measurements of faster spin fluctuations during UDM are currently limited by the maximum fluence that can be applied to a sample in single-shot mode in SAXS~\cite{2010Gutt} or X-ray Fourier transform holography~\cite{2012Wang}. Little is known about the transient magnetic entropy $S_{M}(\tau)$ during UDM. In contrast to $\langle \Delta m(\tau) \rangle$, related to magnetization $\sigma_M$ and field $\langle \zeta_{i} \zeta_{j} \rangle$ fluctuations, magnetic entropy is related to the statistical correlation $\langle \Delta M \Delta S_{M} \rangle$ between magnetization and entropy and a different aspect of ultrafast magnetization dynamics. Experimental access to high equilibrium temperatures opens new opportunities and may be applied to examine the time-dependence of non-equilibrium laser-induced magnetic entropy during ultrafast demagnetization.

\section{Conclusion}

Heat accumulation in nanostructures facilitates measurements at large equilibrium temperatures. Ultrafast demagnetization in Co/Pd ferromagnetic superlattices is measured in new experimental conditions, for large thermal spin fluctuations and small laser-induced transients. The transient demagnetization evolves in parallel with equilibrium thermal demagnetization up to the Curie temperature. Measurements confirm that full demagnetization occurs before all-optical switching in Co/Pd in our experimental conditions. The magnetization time-dependence is fit well with the spin-flip scattering model.

\section*{Acknowledgments}

This research was supported by the University of Louisville Research Foundation. We thank X Wang and J Abersold for assistance with sample preparation at the University of Louisville cleanroom.


\section*{References}

\begin{thebibliography}{}

\bibitem{2010Koopmans}
Koopmans B, Malinowski G, Dalla Longa F, Steiauf D, Fahnle M, Roth T, Cinchetti M and Aeschlimann M 2010 \emph{Explaining the paradoxical diversity of ultrafast laser-induced demagnetization}, Nat. Materials \textbf{9}, 259

\bibitem{2017Bierbrauer}
Bierbrauer U, Weber S T, Schummer D, Barkowski M, Mahro A-K, Mathias S, Schneider H-C, Stadtmuller B, Aeschlimann M and Rethfeld B 2017 \emph{Ultrafast magnetization dynamics in Nickel: impact of pump photon energy}, J. Phys.: Condens. Matter \textbf{29}, 244002

\bibitem{2017Bobowski}
Bobowski K, Gleich M, Pontius N, Schussler-Langeheine C, Trabant C, Wietstruk M, Frietsch B and Weinelt M 2017 \emph{Influence of the pump pulse wavelength on the ultrafast demagnetization of Gd(0001) thin films}, J. Phys.: Condens. Matter \textbf{29}, 234003

\bibitem{2014Gunther}
Gunther S, Spezzani C, Ciprian R, Grazioli C, Ressel B, Coreno M, Poletto L, Miotti P, Sacchi M, Panaccione G, Uhlir V, Fullerton E E, De Ninno G and Back Ch H 2014 \emph{Testing spin-flip scattering as a possible mechanism of ultrafast demagnetization in ordered magnetic alloys}, Phys. Rev. B \textbf{90}, 180407(R)

\bibitem{2014Mendil}
Mendil J, Nieves P, Chubykalo-Fesenko O, Walowski J, Santos T, Pisana S and Munzenberg M 2014 \emph{Resolving the role of femtosecond heated electrons in ultrafast spin dynamics}, Sci. Reports \textbf{4}, 3980

\bibitem{2014Kuiper}
Kuiper K C, Roth T, Schellekens A J, Schmitt O, Koopmans B, Cinchetti M and Aeschlimann M 2014 \emph{Spin-orbit enhanced demagnetization rate in Co/Pt-multilayers}, Appl. Phys. Lett. \textbf{105}, 202402

\bibitem{2017Medapalli}
Medapalli R, Afanasiev D, Kim D K, Quessab Y, Manna S, Montoya S A, Kirilyuk A, Rasing Th, Kimel A V and Fullerton E E 2016 \emph{Multiscale dynamics of helicity-dependent all-optical magnetization reversal
in ferromagnetic Co/Pt multilayers}, Phys. Rev. B \textbf{96}, 224421 (2017)

\bibitem{2008Kazantseva}
Kazantseva N, Hinzke D, Nowak U, Chantrell R W, Atxitia U and Chubykalo-Fesenko O 2008 \emph{Towards multiscale modeling of magnetic materials: Simulations of FePt}, Phys. Rev. B \textbf{77}, 184428

\bibitem{2009Kazantseva}
Kazantseva N, Hinzke D, Chantrell R W and Nowak U 2009 \emph{Linear and elliptical magnetization reversal close to the Curie temperature} Europhys. Lett.  \textbf{86}, 27006

\bibitem{2005Koopmans}
Koopmans B, Ruigrok J J M, Dalla Longa F and de Jonge W J M 2005 \emph{Unifying Ultrafast Magnetization Dynamics}, Phys. Rev. Lett. \textbf{95}, 267207

\bibitem{2014Nieves}
Nieves P, Serantes D, Atxitia U Chubykalo-Fesenko O 2014 \emph{Quantum Landau-Lifshitz-Bloch equation and its comparison with the classical case}, Phys. Rev. B \textbf{90}, 104428 (2014)

\bibitem{2010Battiato}
Battiato M, Carva K and Oppeneer P M 2010 \emph{Superdiffusive Spin Transport as a Mechanism of Ultrafast Demagnetization}, Phys. Rev. Lett. \textbf{105}, 027203

\bibitem{2016Salvatella}
Salvatella G, Gort R, Buhlmann K, Daster S, Vaterlaus A and Acremann Y 2016 \emph{Ultrafast demagnetization by hot electrons: Diffusion or super-diffusion?}, Str. Dynamics \textbf{3}, 055101

\bibitem{2016Bergeard}
Bergeard N, Hehn M, Mangin S, Lengaigne G, Montaigne F, Lalieu M L M, Koopmans B and Malinowski G 2016 \emph{Hot-Electron-Induced Ultrafast Demagnetization in Co/Pt Multilayers}, Phys. Rev. Lett. \textbf{117}, 147203

\bibitem{2017Eschenlohr}
Eschenlohr A, Persichetti L, Kachel T, Gabureac M, Gambardella P and and Stamm C 2017 \emph{Spin currents during ultrafast demagnetization of ferromagnetic bilayers}, J. Phys.: Condens. Matter \textbf{29}, 384002

\bibitem{2015Schmising}
Schmising C, Giovannella M, Weder D, Schaffert S, Webb J L and Eisebitt S 2015 \emph{Nonlocal ultrafast demagnetization dynamics of Co/Pt multilayers by optical field enhancement}, New J. Phys. \textbf{17,} 033047

\bibitem{2012Roth}
Roth T, Schellekens A J, Alebrand S, Schmitt O, Steil D, Koopmans B, Cinchetti M and Aeschlimann M 2012 \emph{Temperature Dependence of Laser-Induced Demagnetization in Ni: A Key for Identifying the Underlying Mechanism}, Phys. Rev. X \textbf{2}, 021006

\bibitem{2017F-JAP}
Hoveyda F, Adnani M and Smadici S 2017 \emph{Heat diffusion in magnetic superlattices on glass substrates}, Journal of Applied Phys. \textbf{122}, 184304

\bibitem{2017F}
Hoveyda F, Hohenstein E and Smadici S 2017 \emph{Heat accumulation and all-optical switching by domain wall motion in Co/Pd superlattices} J. Phys.: Condens. Matter \textbf{29}, 225801

\bibitem{2011Liu}
Liu Z, Brandt R, Hellwig O, Florez S, Thomson T, Terris B and Schmidt H 2011 \emph{Thickness dependent magnetization dynamics of perpendicular anisotropy Co/Pd multilayer films}, J. Magn. Mag. Mat. \textbf{323}, 1623

\bibitem{2011Pal}
Pal S, Rana B, Hellwig O, Thomson T and Barman A 2011 \emph{Tunable magnonic frequency and damping in $[Co/Pd]_{8}$ multilayers with variable Co layer thickness}, Appl. Phys. Lett. \textbf{98}, 082501

\bibitem{2003Koopmans-book}
Koopmans B 2003 \emph{Laser-Induced Magnetization Dynamics}, Spin Dynamics in Confined Magnetic Structures II, Topics Appl. Phys. 87, 253, (Eds. B. Hillebrands, K. Ounadjela)

\bibitem{2005Bigot}
Bigot J-Y, Vomir M, Andrade L H F and Beaurepaire E 2005 \emph{Ultrafast magnetization dynamics in ferromagnetic cobalt: The role of the anisotropy}, Chemical Physics \textbf{318}, 137

\bibitem{2010Boeglin}
Boeglin C, Beaurepaire E, Halte V, Lopez-Flores V, Stamm C, Pontius N, Durr H A and Bigot J-Y 2010 \emph{Distinguishing the ultrafast dynamics of spin and orbital moments in solids}, Nature \textbf{465}, 458

\bibitem{2016Vodungbo}
Vodungbo B, Tudu B, Perron J, Delaunay R, Muller L, Berntsen M H, Grubel G, Malinowski G, Weier C, Gautier J, Lambert G, Zeitoun P, Gutt C, Jal E, Reid A H, Granitzka P W, Jaouen N, Dakovski G L, Moeller S,  Minitti M P, Mitra A, Carron S, Pfau B, Schmising C, Schneider M, Eisebitt S and Luning J 2016 \emph{Indirect excitation of ultrafast demagnetization}, Sci. Reports \textbf{6}, 18970

\bibitem{2014Schmising}
Schmising C, Pfau B, Schneider M, Gunther C M, Giovannella M, Perron J, Vodungbo B, Muller L, Capotondi F, Pedersoli E, Mahne N, Luning J and Eisebitt S 2014 \emph{Imaging Ultrafast Demagnetization Dynamics after a Spatially
Localized Optical Excitation}, Phys. Rev. Lett. \textbf{112}, 217203

\bibitem{2012Vodungbo}
Vodungbo B \emph{et al.} 2012 \emph{Laser-induced ultrafast demagnetization in the presence of a nanoscale magnetic domain network} Nat. Comm. \textbf{3}, 999

\bibitem{2016Chen}
Chen J-Y, Zhu J., Zhang D, Lattery D M, Li M, Wang J-P and Wang X 2016 \emph{Time-Resolved Magneto-Optical Kerr Effect of Magnetic Thin Films for Ultrafast Thermal Characterization}, J. Phys. Chem. Lett. \textbf{7}, 2328

\bibitem{2018F}
Hoveyda F, Hohenstein E, Judge R and Smadici S 2018 \emph{Demagnetizing fields in all-optical switching} J. Phys.: Condens. Matter \textbf{30}, 035801

\bibitem{1963Brown}
Brown W F 1963 \emph{Thermal Fluctuations of a Single-Domain Particle}, Phys. Rev. \textbf{130}, 1677

\bibitem{1970Kubo}
Kubo R and Hashitsume N 1970 \emph{Brownian Motion of Spins}, Supplement of the Progress of Theoretical Physics \textbf{46}, 210

\bibitem{1998Garcia}
Garcia-Palacios J L and Lazaro F J 1998 \emph{Langevin-dynamics study of the dynamical properties of small magnetic particles} Phys. Rev. B \textbf{58}, 14937

\bibitem{2012Atxitia}
Atxitia U, Nieves P and Chubykalo-Fesenko O 2012 \emph{Landau-Lifshitz-Bloch equation for ferrimagnetic materials}, Phys. Rev. B \textbf{86}, 104414

\bibitem{1997Garanin}
Garanin D A 1997 \emph{Fokker-Planck and Landau-Lifshitz-Bloch equations for classical ferromagnets} Phys. Rev. B \textbf{55}, 3050

\bibitem{2014Evans}
Evans R F L, Fan W J, Chureemart P, Ostler T A, Ellis M O A and Chantrell R W 2014 \emph{Atomistic spin model simulations of magnetic nanomaterials}, J. Phys.: Condens. Matter \textbf{26}, 103202

\bibitem{2015Hinzke}
Hinzke D, Atxitia U, Carva K, Nieves P, Chubykalo-Fesenko O, Oppeneer P M and Nowak U 2015 \emph{Multiscale modeling of ultrafast element-specific magnetization dynamics of ferromagnetic alloys}, Phys. Rev. B \textbf{92}, 054412

\bibitem{2017Atxitia}
Atxitia U, Hinzke D and Nowak U 2017 \emph{Fundamentals and applications of the Landau–Lifshitz–Bloch equation}, J. Phys. D: Appl. Phys. \textbf{50}, 033003

\bibitem{1987Leggett}
Leggett A J, Chakravarty S, Dorsey A T, Fisher M A P, Garg A and Zwerger W 1987 \emph{Dynamics of the dissipative two-state system}, Rev. Mod. Phys. \textbf{59}, 1

\bibitem{1974Nitzan}
Nitzan A and Silbey R J 1974 \emph{Relaxation in simple quantum systems}, Journal of Chemical Physics \textbf{60}, 4070

\bibitem{1997Wernsdorfer}
Wernsdorfer W, Bonet Orozco E, Hasselbach K, Benoit A, Barbara B, Demoncy N, Loiseau A, Pascard H and Mailly D 1997 \emph{Experimental Evidence of the Neel-Brown Model of Magnetization Reversal}, Phys. Rev. Lett. \textbf{78}, 1791

\bibitem{2010Gutt}
Gutt C \emph{et al.} 2010 \emph{Single-pulse resonant magnetic scattering using a soft x-ray free-electron laser} Phys. Rev. B \textbf{81}, 100401(R)

\bibitem{2012Wang}
Wang T \emph{et al.} 2012 \emph{Femtosecond Single-Shot Imaging of Nanoscale Ferromagnetic Order in Co/Pd Multilayers Using Resonant X-Ray Holography} Phys. Rev. Lett. \textbf{108}, 267403

\end{thebibliography}
\end{document}